\documentclass[twocolumn]{article}

\usepackage{graphicx}

\begin{document}

\title{Secure and Verifiable Electronic Voting in Practice: the use of vVote in the Victorian State Election}

\author{Craig Burton$^1$, Chris Culnane$^2$ and Steve Schneider$^2$ \\ \small \ $^1$ Victorian Electoral Commission, Victoria, Australia \\ \small \ $^2$ University of Surrey, UK}


\maketitle


\begin{abstract}
The November 2014 Australian State of Victoria election was the first statutory political election worldwide at State level which deployed an end-to-end verifiable electronic voting system in polling places.  This was the first time blind voters have been able to cast a fully secret ballot in a verifiable way, and the first time a verifiable voting system has been used to collect remote votes in a political election.  The code is open source, and the output from the election is verifiable.  The system took 1121 votes from these particular groups, an increase on 2010 and with fewer polling places.
\end{abstract}

\section{Introduction}

Proposals for verifiable electronic voting that provide assurances for secrecy of the ballot and integrity of the election have been in the academic literature since the early 2000's \cite{neff01:e-vote,DBLP:journals/ieeesp/Chaum04,chaum05:e-vote,BenalohSimple06,VoteBox}, but the challenge of making them usable and practical for ordinary voters and integrating them into well-understood election processes has meant that their practical deployment has been slow in coming.  


The State of Victoria has a proud history of innovation in voting systems, having introduced the secret ballot and deployed the world's first secret ballot election in 1856 \cite{McKenna} with government printed ballots and controlled polling place supervision.  More recently, the Victorian Electoral Commission (VEC) was an early adopter of electronic voting, and fielded systems in 2006 and 2010.   The primary driver for electronic voting in Victoria is to provide better accessibility for blind, partially sighted, and motor impaired voters through customised interfaces, and to provide access to voters in languages other than English.  It also provided the opportunity for voters to vote electronically out of state and internationally, enabling more rapid return of the votes into the tallying process.  In Australian elections attendance is compulsory, and this also demands that all efforts must be made by the Election Authorities to enable people to vote.  Consideration of security and transparency issues on this high-consequence system motivated them towards a supervised verifiable voting solution to be offered in polling places.  Australia has no e-voting standards or guidelines (except the Telephone Voting Standards TVS2.0), so the Victorian Electoral Commission were guided by the Voluntary Voting System Guidelines \cite{VVSG} as these were considered to be the most recent, progressive, and considered networked IT security threats.  In particular the system was designed to provide {\em software independence} \cite{rivest2008notion}: that ``an undetected change or error in the software cannot cause an undetectable change or error in the election outcome''.

\begin{figure*}
\begin{center}
\includegraphics[width=0.75\linewidth]{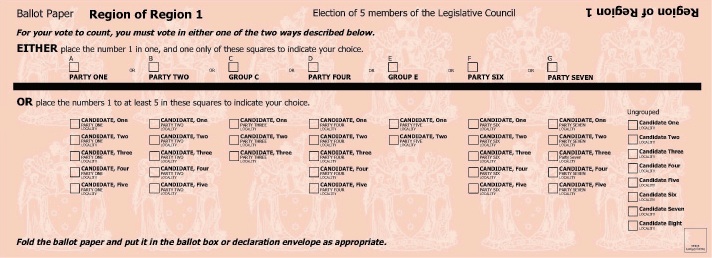}
\hspace*{1cm}
\includegraphics[width=0.125\linewidth]{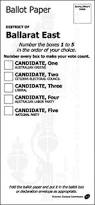}
\end{center}
\caption{Legislative Council and Legislative Assembly Ballot forms}
\label{fig:ballot}
\end{figure*}

The election system in Victoria poses particular challenges for any verifiable solution, because of the complexity of the ballot.  Voters vote in two races in State Elections, using ballot forms illustrated in Figure~\ref{fig:ballot}.  For the Legislative Council voters are offered a list of around 40 candidates to rank in their preferred order, listed Below The Line (BTL).  They may instead select a preference order provided by a party by selecting one choice Above The Line (ATL).  For the Legislative Assembly voters rank all of the candidates in preferential order. 

There are 8 regions comprising 88 districts, so there are 96 races in total.  Nominations for candidates were open until Noon on Friday 14th November 2014, and the electronic system needed to be ready to take votes from Monday 17th November.  There is a period of two weeks of ``early voting'' for which the electronic system is deployed, leading up to the official election day on Saturday 29th November.  Electronic voting was only available during early voting.   Voters are allowed to use any polling station anywhere to cast a ballot in their home race.  Thus all polling stations must offer ballot forms for all the races across the State.  This is also a requirement for the electronic system.

The total number of registered voters for the 2014
election was 3.8 million, of whom 2014 Australian Bureau of Statistics
\cite{abs} indicate as many as 186,000 Victorian travellers,
100,000 adults not proficient in English
and 118,000 adults with low vision or blindness\footnote{Blind Citizens Australia, Australian Blind and Vision Impaired Statistics\\
} were eligible to vote using the electronic system.  Note also that in 2009 7.8\% of 3.8m (8.6\% minus Diabetics as main low vision cohort) needed help with reading or writing, making a further 296,000 additional to the numbers above.

A total of 1121 votes were collected.   This was more votes than were collected by the 2010 electronic system, and the system was deployed at fewer locations.



\section{Related work}

The only end-to-end verifiable political elections to date have taken place in Takoma Park, Maryland, US, where the Scantegrity system was successfully used in 2009 and 2011 in the municipal election for mayor and city council members \cite{SIITakPk}.   This groundbreaking work demonstrated the feasibility of running an election in a verifiable way.  However, the Scantegrity system becomes impractical with a preferential ballot with 40 candidates, and would require non-trivial changes to its design to handle a state-wide election.

Other verifiable systems such as Helios \cite{DBLP:conf/uss/Adida08} and Wombat \cite{wombat2} have been used for Student Union and other non-political elections, but scaling them up to politically binding elections and hardening them for a more robust environment is challenging, and in their present form they are also not so suitable for elections as complex as those in Victoria.  Other electronic elections that have been fielded (such as those recently fielded in Estonia, Norway, Canada, and New South Wales) do not provide end-to-end verifiability.  Some of them allow voters to verify some aspect of their vote (for example, that the system has a vote with a particular receipt number) but do not provide verifiability all the way from the vote cast through to the final tally.

\section{System Description}

The starting point for the design of the system was the Pr\^et \`a Voter split ballot \cite{chaum05:e-vote}, a printed ballot form with the candidates listed in a randomised order (different on different ballot forms), where the voter makes their selection and then separates and destroys the list of candidates, retaining and casting the marked preference list for verifiable tallying.  Numerous additional design steps were required to make this system practical in the context of the Victorian election, and the development of the design lasted over a period of about two years.  The completed design is described in \cite{CRSTArXiV14}.   The first critical development was to adapt the design to incorporate electronic assistance to capture the vote (necessary for such a large number of candidates), so an Electronic Ballot Marker was introduced: a tablet that provided a voter interface for capturing the vote.  The design was also adapted to handle preferential voting; the ballot forms needed to be printed on demand in the polling places since voters can vote in any race for which they are registered; a secure and robust distributed web bulletin board needed to be designed and implemented for accepting the votes and verifiably making the information public; and some technical solutions were required for compressing votes for processing them, to cope with the large number of candidates.  


\subsection*{End-to-end verifiability}

The end-to-end verifiability provided by the system relies on a number of checks, including some which must be performed by some of the voters.   Once a voter is validated as being eligible to vote, they are provided with a candidate list (CL) with the names in a random order.  

Voters choose either to use this list to cast their vote, or to have it audited.  Voters are allowed to demand that the list they are given is `audited' to check that the printed order matches the encrypted order that the system has already committed to.  An audited CL is decrypted and so cannot then be used to vote, since it would expose the vote.  Following audit a voter would need to be issued with a new CL.  In practice, for this initial deployment voters were not alerted to this possibility since there was a concern that the subtlety of what it was achieving (essentially random sampling of correct construction of the ballot forms) was too complex for voters to absorb on the spot.  For future deployments some advance education would raise awareness of this step.

\begin{figure}
\begin{center}
\includegraphics[width=0.9\linewidth]{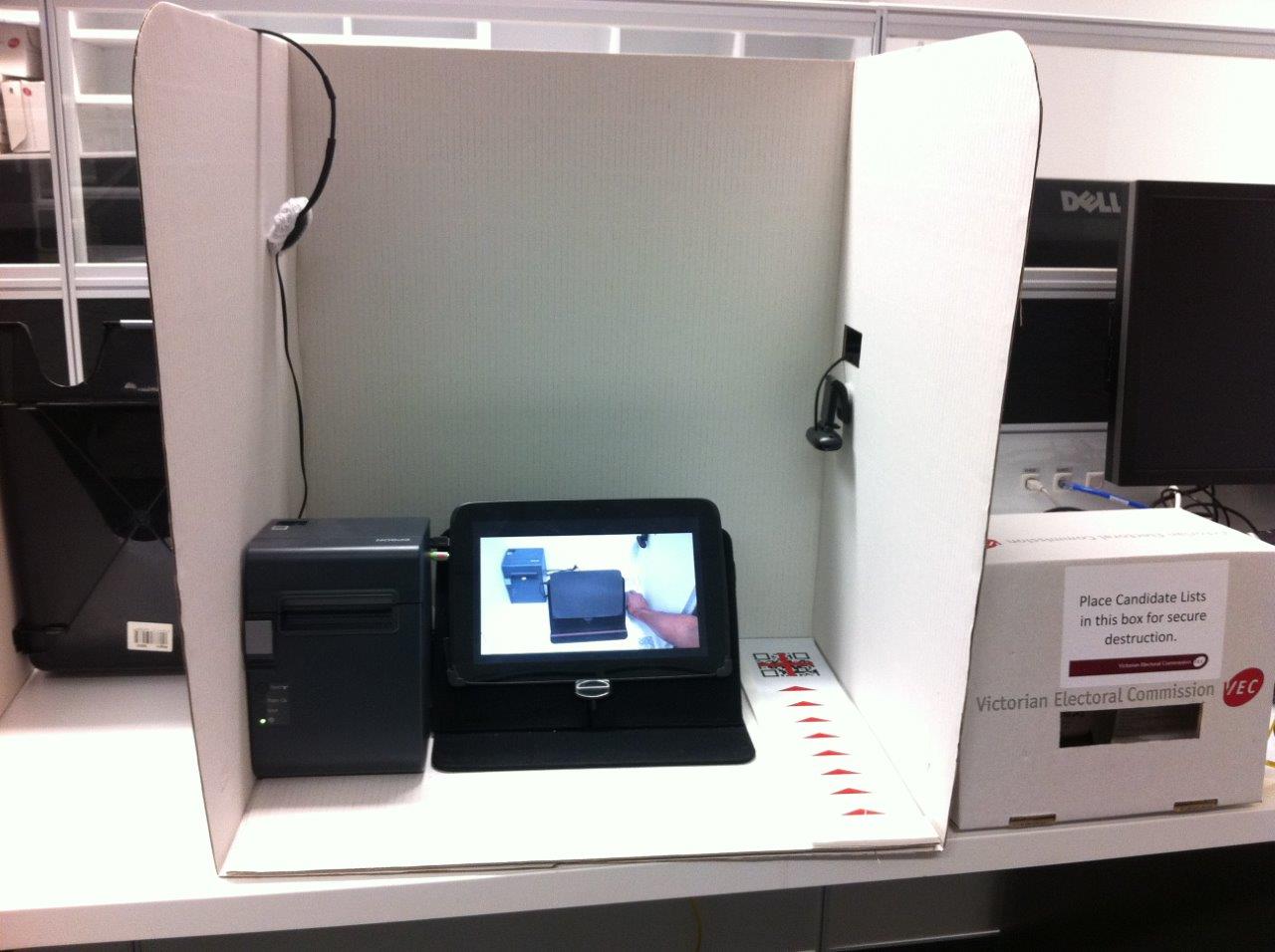}
\end{center}
\caption{A voting booth} 
\label{fig:booth}
\end{figure}

\begin{figure}
\begin{center}
\includegraphics[width=0.9\linewidth]{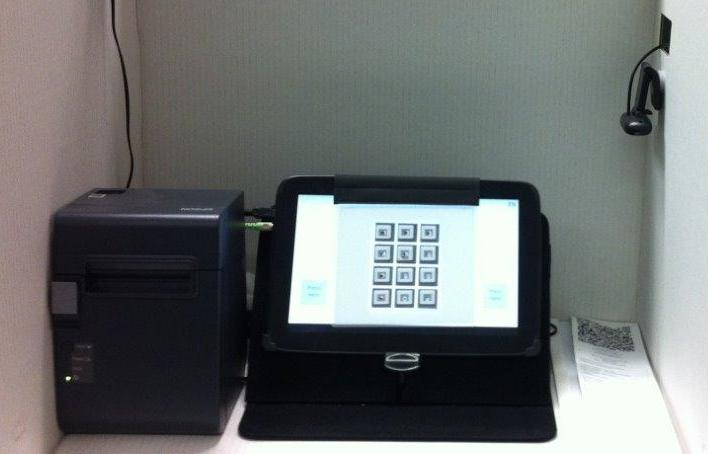}
\end{center}
\caption{A voting booth for use by blind voters.  Note the tactile `telephone keypad' overlay over the touchscreen, and the headphones for audio instructions} 
\label{fig:blindvoters}
\end{figure}

\begin{figure}
\begin{center}
\includegraphics[width=0.9\linewidth]{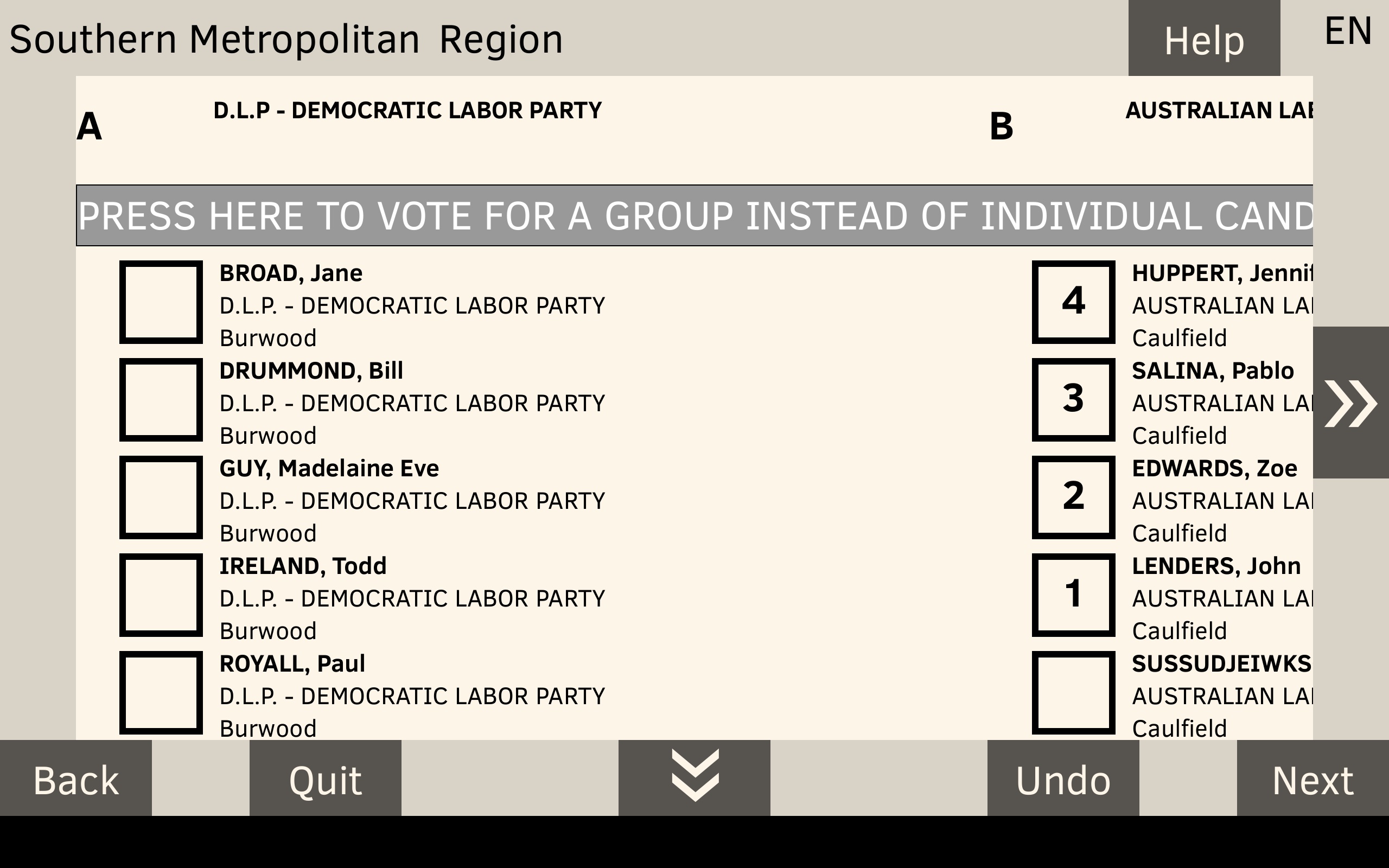}
\end{center}
\caption{Tablet interface for capturing votes} 
\label{fig:interface}
\end{figure}

\begin{figure}
\begin{center}
\includegraphics[width=0.78\linewidth]{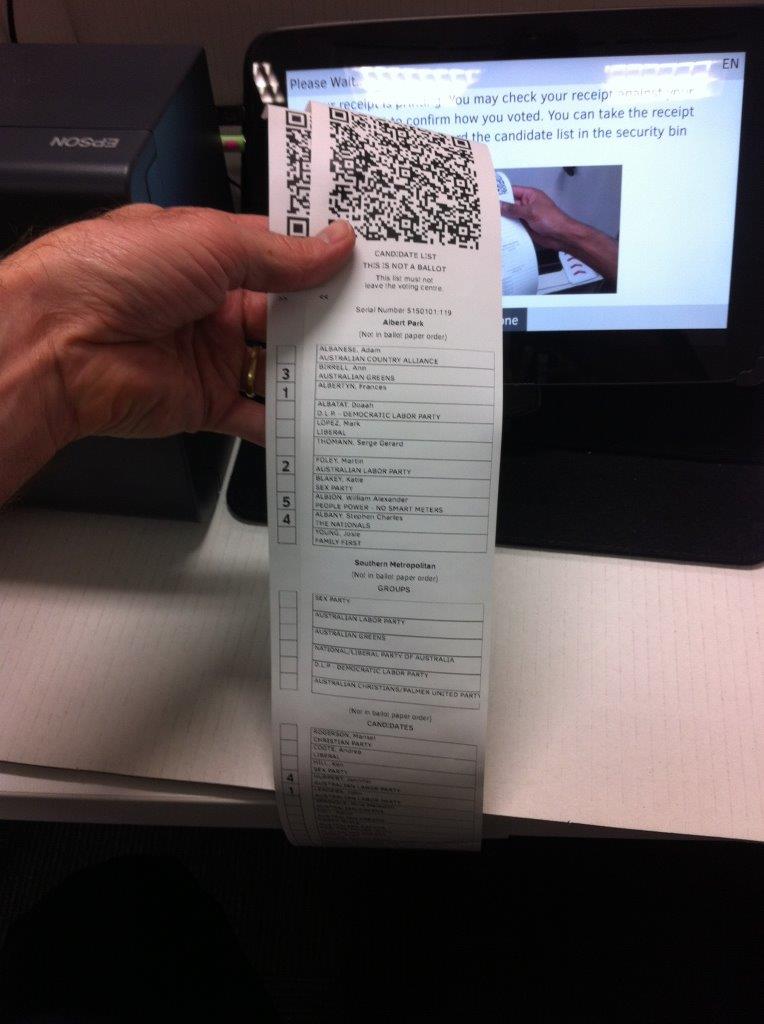}
\end{center}
\caption{Matching the candidate list and preference receipt} 
\label{fig:CLPR}
\end{figure}

If the voter instead uses the list to vote, then it is scanned into the tablet.  The booth setup is illustrated in Figure~\ref{fig:booth}, and the setup for blind voters is illustrated in Figure~\ref{fig:blindvoters}.  Scanning of the candidate list QR code launches the vote capture application, which allows the voter to enter their vote via the telephone keypad overlay.  The regular tablet interface for sighted voters is illustrated in Figure~\ref{fig:interface}.  Having voted, a preferences receipt (PR) with the preferences in the same order is printed separately.  The voter verifies that the preferences match the correct candidates, by comparing the lists side by side, as in Figure~\ref{fig:CLPR}.  This check ensures that the receipt captures the vote as cast.  Once this is done the candidate list must be destroyed in order to keep the vote secret.  The PR is retained by the voter.  The fact that the candidate names were in a random order ensures that the PR does not expose who the preferences were for and thus provides ballot secrecy.  

The system allows voters to `quarantine' or cancel their vote if the PR is not provided for any reason, or if the voter considers it to be incorrect, and in such cases the voter may cast their vote again.

The voter later looks up their PR on the Web Bulletin Board, and verifies that it has been included properly.

It is important to add that all aspects of the voting interfaces as well as the verification measures intended for electors were also made accessible.  A ballot audit could be taken to an EBM and the device would read the contents out.  Both the CL and PR could be read out separately on any EBM or if provided together, the assembled preference-order vote could be read back to the elector.  All EBMs were provided with headphones and the voting interfaces provided three completely separate forms for those with some vision; for those who could use touch-screen gestures; and, for those who wanted a telephone-IVR type interaction - which was provided via a latex screen overlay giving the tactile impression of the 12 telephone buttons.

The system deployment also made use of conventional security features of course, such as how it was remote managed, logging, and so on.  

End-to-end verifiability is achieved through the collection of all the steps above.  Voters can check that their vote is {\em cast as intended}, since they check the list of candidates against the preference list on their receipt.  They confirm their vote is {\em recorded as cast} by checking that their preferences on the receipt correspond to the information recorded on the bulletin board.  Finally, it is publicly verifiable that the correct tally is returned, {\em counted as recorded}, because the mixing and decrypting of the encrypted votes generates cryptographic proofs that can be independently checked.  This provides a chain of links all the way from the initial creation of blank ballots to casting of the vote right through to the tallying.

In this election the electronic votes needed to be combined with paper votes, and so the electronic votes were printed off to be included within the count of the paper votes.  Printed vote also bore a file line number which could be used to check any printed e-vote with the emitted verifiable clear text votes online.  In this case we have verifiability through to the decrypted votes: that any cast vote did make it into the paper count, which was then done in the usual way.


%
%

The front-end user interface and election administration was developed in-house by VEC.  VEC contracted the University of Surrey to develop the back end, SuVote \cite{suvote}, which was responsible for creating ballot forms, accepting the votes and managing the web bulletin board.  VEC contracted Cryptoworkshop.com to develop the Mixnet, Ximix \cite{ximix}.  The project also involved a wider academic advisory team from the Universities of Surrey, Melbourne and Luxembourg.  The independent review of the design and code \cite{demtechreview} was conducted by Demtech.

\section{Deployment}

Because this was a completely novel system, VEC rolled out in a limited deployment in 24 early voting centres around Victoria including 6 ``accessibility super centres'', and offered only to the particular legislated groups of voters.  It was also deployed in the Australia Centre, London, UK where legislation made it available to all voters who were casting their vote from there in order to gain experience of the remote voting solution.

The total number of votes that were received over the two weeks were 1121, of which 973 were from London and the remaining 148 from the 24 centres in the State of Victoria.   

In fact the system was developed for much higher demands in order to scale up for future elections:  it handled 1 million votes in testing, and under stress was able to respond to individual voters within 10s, and to accept 800 votes in a 10s period.  

\section{Outcomes}
A range of instruments were used to evaluate this project.  A University of Surrey survey of voters leaving the Australia Centre in London having cast their votes is most indicative of the system with the entire voter cohort.  For Victoria, VEC ran an anonymous online questionnaire of the poll workers asked questions about equipment setup and voter support.  Both surveys asked about verifiability, trust and security.  VEC also ran its own survey of London electors as well as Victorian electors.

To analyse time-to-vote, receipt lookup and other measurements, server logs and web server analytics were used.  Google Analytics collected information for public-facing information and lookup services.  It should be noted that the voting protocol does not capture voting interface actions and that the Ballot Marker is stateless.  These privacy controls prevented more conventional usability data being captured such as user actions on the ballot faces.  Prior to the live election the system was instead usability tested by VEC with several cohorts of voters with barriers and with non-English speaking (non-bilingual) volunteers.  The system usability was assessed by an expert third party organisation.

\subsection*{Voter surveys}

The London data has a very low (1.6\%) non-answer rate and responses were collected by approaching electors at random.  The survey form is attached in Appendix~\ref{app:survey}.  Some questions were taken from an international benchmark\cite{DBLP:conf/stast/KarayumakKOVV11} to provide a level field for measuring verifiability effectiveness via electors.

\begin{figure}
\begin{center}
\includegraphics[width=0.9\linewidth]{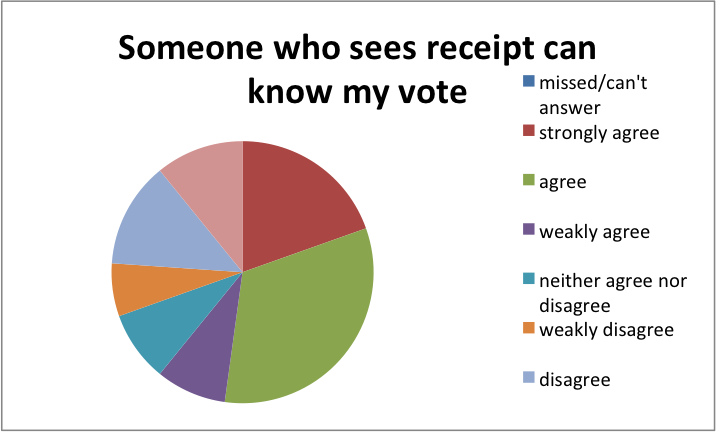}
\end{center}
\caption{Responses to the statement: ``Someone who sees your receipt can know your vote''} 
\label{fig:shows}
\end{figure}

The overall results were that voters were generally satisfied with the usability of the system, but there was a wide variation in understanding of the security assurances provided.  For example, some voters answered that the receipt showed their vote (which it does not, since a receipt should never link directly to any vote).  The results are shown in Figure~\ref{fig:shows}.  Although most voters trusted the system implicitly they nonetheless took part in the verifiability steps and many said they would check receipts at home.  The resistance to such new steps reported in \cite{DBLP:conf/stast/KarayumakKOVV11}  was not observed in London.  No voters reported that the process took``too long'' despite the added task of comparing CL to PR. 

In more detail, the headline results from the {\bf voter surveys} were as follows:
\begin{enumerate}
\item	Respondents found the system easy to use, as illustrated in Figure~\ref{fig:easy}.   75\% or greater respondents stated Agree or Strongly Agree to all positive aspects of usability.  75\% stated they preferred the system to paper voting.  This is also evident in comments and in the time taken to vote.  60\% of respondents voted in 4 minutes or under, with 96\% in 5-10 minutes or less.  This was the whole time to vote on the system (not the time to get a PR or wait in queues).  None stated the process took ``too long''.  87\% stated Agree or Strongly Agree to the statement ``I would tell other people to use this system,'' illustrated in Figure~\ref{fig:tellothers}.
\begin{figure}
\begin{center}
\includegraphics[width=0.7\linewidth]{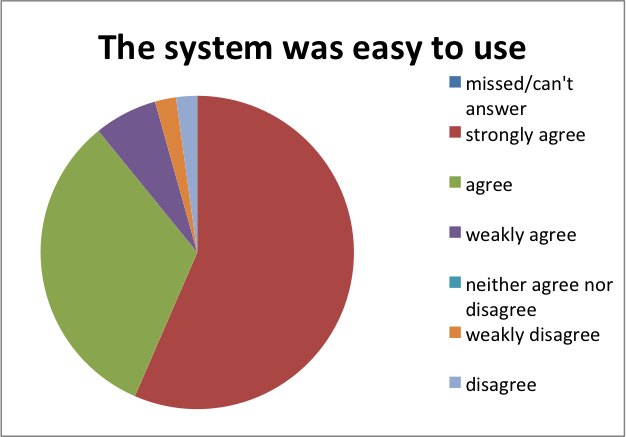}
\end{center}
\caption{Responses to the statement: ``The system was easy to use''} 
\label{fig:easy}
\end{figure}
\begin{figure}
\begin{center}
\includegraphics[width=0.7\linewidth]{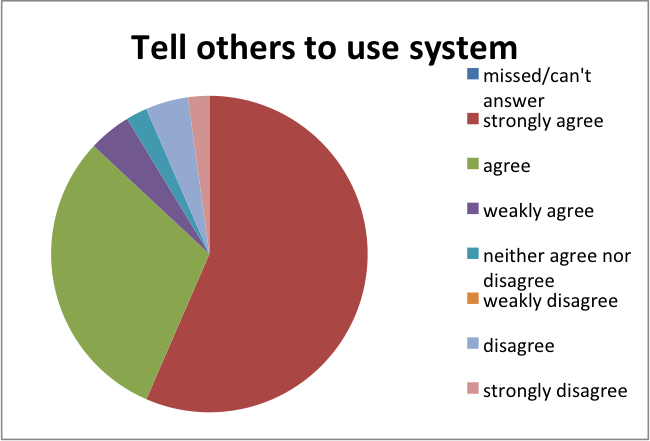}
\end{center}
\caption{Responses to the statement: ``I would tell other people to use this system''} 
\label{fig:tellothers}
\end{figure}
\item Respondents trusted the voting system, and this correlated strongly with the user having a good user experience.  Just over 60\% of respondents had no concerns regarding e-voting security.
\item Respondents found the verification lists easy to use.  About half of the respondents compared the CL and PR together.
This rate of checks provides a confidence level of over 99.9\% that votes were not changed on submission to the Bulletin Board in sufficient numbers to change the outcome.  The closest race was Prahran with a winning margin of 41 votes, meaning that a minimum of 21 would need to have been altered (from the reported runner-up to the reported winner, in the unlikely event that the reported runner up in fact won by a single vote) in order to have changed the outcome in that race.   
\item  About 40\% stated they were Very Likely or Likely to verify their receipt on the VEC website.  In fact there were around 150 receipt lookups (the figure of 250 in Table~\ref{fig:webhits} includes just under 100 tests), about 13\% of the electronic votes cast.   
\item Many respondents did not understand the purpose of the verification measures.  This is evident in low correlation with questions about trust and security as well as from comments.  More than half of respondents thought that the voting receipt gave away the content of their vote, which is not the case.  However, for people ``concerned about e-voting'', comprehension of the receipt was much better, only a quarter felt the receipt leaked their vote.
\item One desired outcome of the survey was not observed: that at least some respondents would use verification measures {\em because} they have concerns about e-voting.  That is, the survey could not detect the kind of vigilance that the verification relies on via negative correlations between ``trust'' and ``use of the verification measures''.   
\end{enumerate}

\subsection*{Poll worker surveys}

The {\bf poll worker} surveys were conducted after the end of the election.  There is some over-reporting as may be expected since the survey responses can include the same events reported separately.  About half the staff who supported e-voting responded. The summary of findings is:
\begin{enumerate}
\item System features for accessibility were well used.  A quarter of respondents set font or contrast for users, with forty percent setting audio mode.  Twelve percent of respondents reported setting a non-English language.
\item The system did not require much intervention in the voting session, and when this occurred, the intended support tools were used.  In fewer than ten percent of cases, staff had to complete the e-vote for electors.  A quarter of respondents reported using the {\em switch to visual} support feature to help an audio voter in-session.  No respondents needed to use the {\em switch to English} support feature.
\item The verifiability measures were well used.  A quarter of respondents saw electors perform CL-PR.  Only two respondents handled electors reporting the PR did not match their vote.
\item Staff may have not fully understood verifiability.  Three quarters of respondents stated they Strongly Agree or Agree to understanding verifiability and the printed lists.  However, the same respondents answered differently to questions asking them about the lookup of receipts on the web and of CL audit.    
\item Although more than half of respondents stated the system was Too Difficult to Operate or Not Very Reliable, two thirds stated they would be happy to support it if more voters came to use it.  Two thirds stated they were happy or indifferent to e-voting.
\end{enumerate}

\subsection*{Web lookups}

The vVote suite of pages on vec.vic.gov.au were all visited with increasing frequency up to Election Day.  Pages such as the Electronic Voting page were hit more than 20,000 times, by 18,600 unique viewers.   The totals are given in Figure~\ref{fig:webhits}.
\begin{figure*}
\begin{center}
\begin{tabular}{lllll}
Section/page &	Sub-page	& Go Live	& Views	& Reading Time \\ \hline
Assistance for voters	 &&	1/11	& 11,120	\\ \hline
Electronic voting &&		12/11 &	20,426 &	0:04 \\ 
&	Electronic voting detail	& 22/11	& 408	& 0:15 \\
&	Find receipt / fetch	& 18/11	& 250	\\
&	Verification data files &	18/12 &	139	\\
&	Source code repository &	12/11 &	55	\\
&	Information Files &	1/11	& 1505	\\ \hline
Candidate Registration	& Candidate Name  &	6/11 & 	379	\\
& Pronunciation (to 15/11) 
\end{tabular}
\end{center}
\caption{VEC Electronic Voting web page hits: November 2014 -- January 2015}
\label{fig:webhits}
\end{figure*}


Users accessed support documents such as locations of assistive voting centres (950),  information about electronically assisted voting (554), and the Demtech assessment of vVote (35).  

People accessing the vVote suite of pages came most frequently from LinkedIn (90) and Vision Australia (54) and Twitter (24), among others.  Google search outside of VEC was used to access this suite 1071 times with 24 site-internal searches for `Electronic Voting'.

The rate of receipt lookups (13.4\%) provides confidence at the 95\% level that there have not been sufficient missing, damaged or changed votes to change the outcome, even in the closest race, Prahran\footnote{If 21 votes were changed (the minimum possible to alter a result) and each change has a $86.6\%$ probability that it will not be detected, then the probability that none would be detected is $0.866^{21} = 0.049$} .  This means that even if there was cheating in the recording of votes on the bulletin board, then there would be at least a 95\% chance of detecting this.  Furthermore, additional voters can check their receipts after the election, which would make the confidence level still higher.  In fact there were no complaints about altered or missing votes.    

\subsection*{Uptime}
\begin{enumerate}
\item	The Web Bulletin Board systems were up 100\% with no errors.  Average response (reply) time was 0.3 seconds.  
\item A full analysis of the log files showed that no unexpected exceptions occurred during live voting.
\item London was offline intermittently, totalling about 14 hours of downtime over the two weeks due to networking problems.  Voters affected by this voted on paper ballots.
\end{enumerate}

\subsection*{Process times}
London reported queuing and some network problems and Victoria served electors with a range of barriers and impairments.  
We report on London and on Victoria separately:
\subsubsection*{London}
\begin{enumerate}
\item The average time to process a vVote elector was 275.6 seconds---about 4.5 minutes.
\item The average voting session time (excluding printing the candidate list and then waiting to vote) was 172.2 seconds---about 3 minutes.   
\item The shortest process time was 46 seconds, or which the voting session was 14.6 seconds.  
\item The longest process time was 1089.4 seconds with the voting session taking 100 seconds less.
\item Average time to vote ATL was 152.6 seconds---about 2.5 minutes.
\item Average time to vote BTL was 270 seconds---about 4.5 minutes.
\end{enumerate}

\subsubsection*{Victoria}

\begin{enumerate}
\item The average time to process a vVote elector was 672.8 seconds---about 11 minutes.
\item The average voting session time (excluding printing the candidate list and then waiting to vote) was 570.4 seconds---about 9.5 minutes.   
\item The shortest process time was 40.9 seconds.  The min voting session was 19.8 seconds.  
\item The longest process time was 3895.8 seconds (just over an hour) with the voting session 300 seconds less.
\item Average time to vote ATL was 542.8 seconds---about 9 minutes.
\item Average time to vote BTL was 658.3 seconds---about 11 minutes.
\end{enumerate}

The proportion of formal electronic votes (i.e. ballot forms that were properly completed and hence contain votes that will be counted) was 98.13\%, with 29 informal votes in District races (2.5\%), and 13 informal votes in Region races (1.15\%).
%
%
The main informality was people picking one only candidate in the District race.  Electors receive visual/audio/in-language warnings as well as instructions to complete the ballots but formal voting is not enforced by the system. For the votes taken on paper the formality rate was 95.7\%.  In 2010 the electronic formality rate was 99\%, and for paper votes it was about 95\%.   These statistics support the claim that the electronic interface reduces the number of electors who accidentally spoil their ballot. 

%

\section{Discussion and Lessons learned}

There was an inevitable tension between the desire to allow voters to ``vote and go'' (i.e. to keep the voting experience as lightweight as possible) and the need to have security steps that some of the voters follow, to ensure verifiability.  The voter surveys found that the voters were generally satisfied with their voting experience, so the security elements did not obstruct the voting process for them, and those that wanted to vote and go were able to do so.  Electors did not resist the verifiability features as was observed in a verifiable voting mock election study \cite{DBLP:conf/stast/KarayumakKOVV11}.

A key concern of the vVote project was that the verification system might cause electors to become confused about whether their vote was taken as it was cast.  It was estimated that the Pr\^et \`a Voter voting receipt randomisation of preferences would cause electors who checked them to think their actual vote had been changed.  For this reason, all surveys included questions about this risk.  The results show there were some isolated cases of confusion: four electors in London and two at Victorian sites reporting explaining this at least once; and there was one letter directly to VEC.  In all of these cases the issues were resolved satisfactorily.  It is not the case that electors reported, or staff observed, electors substantially confused about verification receipts.

The system is a verifiable system and its verifiability audits were well used in 2014.  However, one audit mechanism (ballot audit: voters being able to challenge and confirm a candidate list) was not promoted for this initial deployment since there was concern it might confuse voters unfamiliar with the concept of verifiability.  More work is needed for future deployments to make Ballot Audit simple and accessible to both staff and electors so that is it provided in accordance with the system protocol.  

The system architecture was entirely housed at VEC rather than distributed across different locations to increase resilience.  vVote servers should not be housed together, for both disaster recovery reasons and also for the Electoral Commission's plausible deniability in keeping hands off systems that can otherwise collude or be observed by the Commission.  
%




Although the system was developed for use in the State of Victoria, much of it can be customised to elections elsewhere.  The system essentially captures and posts votes in a verifiable way, and then the posted votes can be tallied separately.  This means that the system is not tied to a particular voting scheme, or voting process, and can be customised to handle different schemes involving for example single choice, multiple choice, preference voting, or alternative vote.  The system is open source and it is hoped the findings here and techniques present in the sources lead to greater use of this approach to electronic voting.


\subsection*{Acknowledgements}
We are grateful to Peter Ryan and Vanessa Teague for their support and encouragement during the deployment of the system, and for numerous discussions on all aspects of it.   Thanks to EPSRC for funding the research underpinning this system under grant EP/G025797/1, `Trustworthy Voting Systems', and under the Surrey Impact Acceleration Account EP/K503939/1.  Thanks also to staff at the Victorian Electoral Commission for their support of this project.

\bibliographystyle{ieeetr}
\bibliography{e-vote}

\appendix

\section{University of Surrey London Survey Instrument} 
\label{app:survey}

\begin{figure*}
\includegraphics[width=0.9\linewidth]{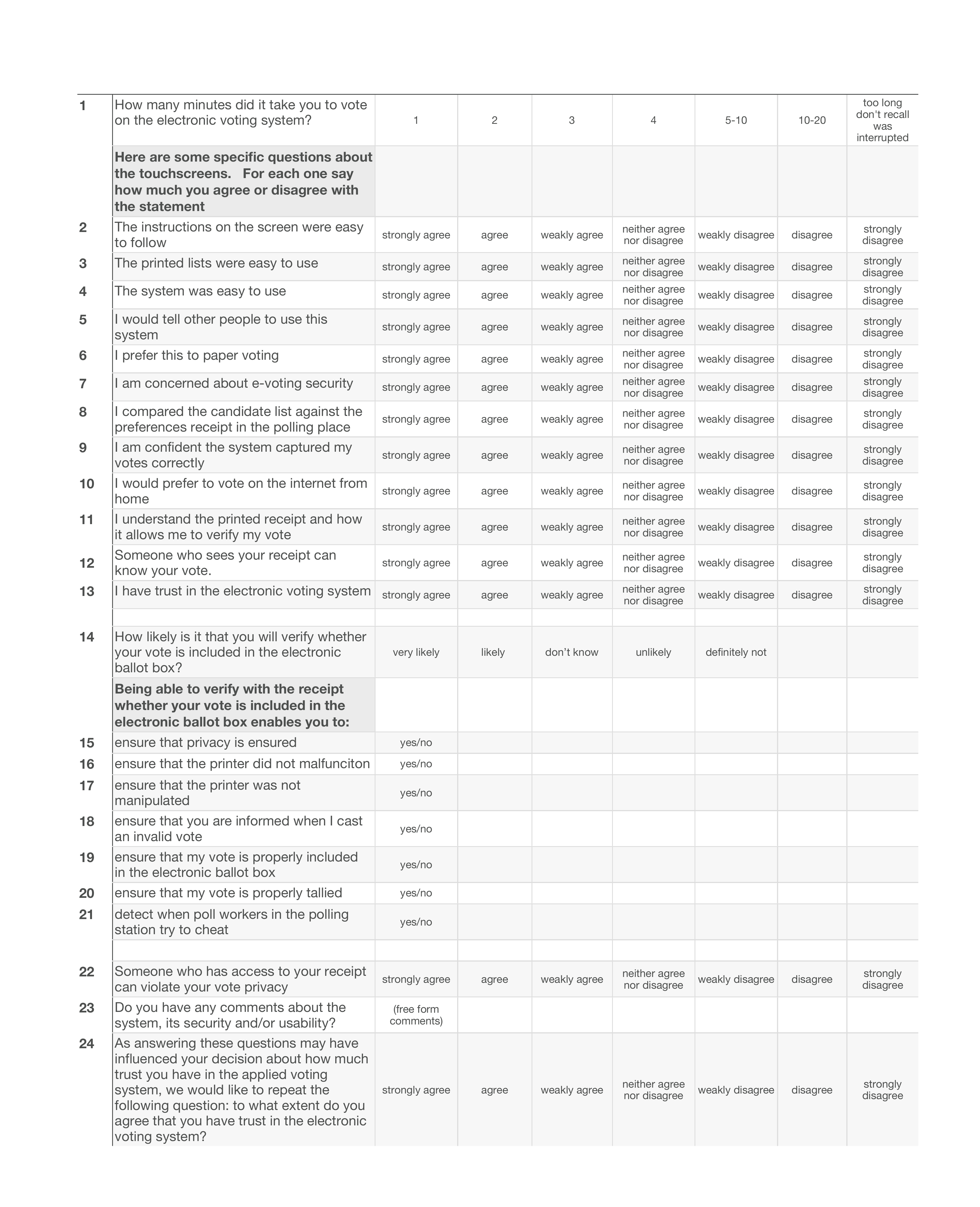}
\caption{London Survey Instrument}
\end{figure*}

\end{document}